\documentclass[%
 reprint,
 amsmath,amssymb,
 aps
]{revtex4-2}

\usepackage{graphicx}
\usepackage{dcolumn}
\usepackage{bm}
\usepackage{float}
\usepackage{multirow}
\usepackage{hyperref}
\usepackage{comment}

\begin{document}

\preprint{APS/123-QED}

\title{Disentangling Flow Contributions from the Chiral Magnetic Effect in U+U Collisions with Forward–Backward Multiplicity Asymmetry}

\author{Kaiser Shafi}
\email{kaisers@iiserbpr.ac.in}
\author{Sandeep Chatterjee}
\email{sandeep@iiserbpr.ac.in}
\affiliation{Indian Institute of Science Education and Research Berhampur}
\date{\today}

\begin{abstract}
The observation of the Chiral Magnetic Effect (CME) in heavy-ion collisions remains challenging because of large flow-induced backgrounds and experimental constraints. We demonstrate that the forward–backward multiplicity asymmetry (FBMA) provides a robust and experimentally accessible control parameter to separate the flow background from CME signal in the collisions of deformed nuclei, such as prolate uranium where FBMA is naturally enhanced and correlated with the initial-state geometry. Monte Carlo Glauber simulations indicate that varying FBMA within a fixed centrality class modulates ellipticity largely independently of the magnetic-field correlator, establishing FBMA as a practical tool for disentangling CME signals from flow driven background.
\end{abstract}

\maketitle

\section{Introduction}
A central objective of contemporary nuclear physics is to understand the properties of Quantum Chromodynamics (QCD) under extreme conditions. Ultrarelativistic heavy-ion collisions (HICs) at the Relativistic Heavy Ion Collider (RHIC) and the Large Hadron Collider (LHC) create a hot and dense state of matter known as the quark--gluon plasma (QGP), in which color degrees of freedom are deconfined and chiral symmetry is partially restored~\cite{Shuryak:2004cy,Gyulassy:2004zy}. Beyond its collective properties, the QGP provides a unique environment for investigating non-perturbative phenomena associated with the topological structure of the QCD vacuum.

Gauge field configurations with nontrivial topology, such as instantons and sphalerons, can induce local violations of parity ($\mathcal{P}$) and charge--parity ($\mathcal{CP}$) symmetries through anomalous interactions that change quark chirality~\cite{Kharzeev:1998kz,Kharzeev:2004ey}. Although these topological transitions are local and short-lived, they may generate observable consequences in final-state particle correlations produced in heavy-ion collisions.

A prominent manifestation of these effects is the chiral magnetic effect (CME), which predicts the generation of an electric current along the direction of an external magnetic field in a chirally imbalanced medium~\cite{Fukushima:2008xe,Kharzeev:2015znc}. In non-central HICs, ultra-strong magnetic fields are produced by the relativistic motion of spectator protons, reaching magnitudes of $eB \sim m_\pi^2$~\cite{Skokov:2009qp,Bzdak:2011yy}. The interplay between these strong magnetic fields and locally $\mathcal{P}$- and $\mathcal{CP}$-odd domains can result in charge-dependent azimuthal correlations, characterized by preferential emission of same-sign charged particles along the magnetic field direction~\cite{Kharzeev:2007jp,Voloshin:2004vk}.

Experimentally, the CME has been investigated using correlation observables sensitive to charge separation relative to the reaction plane~\cite{STAR:2021mii}. One such observable is the three-particle correlator~\cite{Voloshin:2004vk},
\begin{equation}
\gamma^{ab} = \left\langle \cos \left( \phi^a + \phi^b - 2 \psi_{\mathrm{RP}} \right) \right\rangle,
\label{eq:gamma_corr}
\end{equation}
where $\phi^{a,b}$ denote the azimuthal angles of particles with charges $a,b=\pm$, and $\psi_{\mathrm{RP}}$ is the reaction plane angle. This correlator has been measured extensively in Au+Au, Cu+Cu, and Pb+Pb collisions~\cite{STAR:2009wot,STAR:2014uiw,ALICE:2012nhw}. While designed to suppress reaction-plane--independent correlations, its interpretation is complicated by significant background contributions arising from elliptic flow ($v_2$), resonance decays, and other conventional sources that can mimic CME-like signals~\cite{Pratt:2010zn,Bzdak:2009fc}.

A major challenge in isolating the CME arises from the intrinsic correlation between the magnetic field strength and the initial collision geometry. The magnetic field is largely determined by spectator nucleons, while elliptic flow is driven by the initial spatial anisotropy of the overlap region. Consequently, attempts to suppress flow-related backgrounds such as selecting more central collisions also suppresses the magnetic field strength and the expected CME signal~\cite{Voloshin:2010ut}. This interplay motivates the development of alternative strategies that vary the initial geometry while maintaining sensitivity to the magnetic field.

Collisions of deformed nuclei provide a promising avenue in this regard. The uranium nucleus exhibits a prolate deformation, giving rise to a wide range of initial geometries depending on the relative orientations of the nuclear symmetry axes~\cite{Voloshin:2010ut,Haque:2011aa}. Distinct configurations, commonly referred to as Tip--Tip, Body--Body, and Body--Tip, result in markedly different initial eccentricities and magnetic field strengths at comparable impact parameters~\cite{Masui:2009qk,Chatterjee:2014sea}. In particular, Body--Tip configurations in U+U collisions can generate strong magnetic fields due to spectator contributions while exhibiting relatively small spatial anisotropy in the overlap region, making them well suited for CME studies with reduced flow-induced backgrounds~\cite{Chatterjee:2014sea}.

Previous experimental efforts to exploit these geometric variations have often relied on spectator neutron asymmetry, defined as the absolute difference between the number of neutrons detected in forward and backward Zero Degree Calorimeters (ZDCs)~\cite{Wang:2013dza,Chatterjee:2014sea}. Although this observable provides sensitivity to event-by-event geometry fluctuations, it is subject to experimental limitations such as finite acceptance, neutron detection inefficiencies, and resolution effects~\cite{Adler:2000bd,Grachov:2006ke,Xu:2016alq,Zugravel:2025fpm}. Hence in this work, we explore the possibility of a different trigger for these events, namely FBMA as was demonstrated in Ref.~\cite{Bairathi:2015uba}.

\section{Model Description}
We employ the shadowed Monte Carlo Glauber model (shMCGM), a modification of the standard two-component Glauber ansatz in which entropy deposition from a given participant nucleon in the projectile nucleus is suppressed due to the presence of other nucleons in the projectile nucleus that shadow it from the target nucleus and vice-versa for the participants in the target nucleus~\cite{Chatterjee:2015aja}. This shadowing mechanism weakens the simple two-component Glauber model scaling and has been shown to improve the description of ultra-central uranium–uranium collisions, including the absence of the predicted “knee” in the ellipticity–multiplicity correlation observed by STAR~\cite{Chatterjee:2015aja}. 

In this setup, the final pseudorapidity ($\eta$) dependent charged particle multiplicity $\frac{dN_{Ch}}{d\eta} \left(\eta \right)$, proportional to the initial spacetime rapidity ($\eta_s$) dependent entropy density, is calculated as the sum of three terms (Eq.~\ref{eq:dNchdeta_eta}). These terms arise from the participant nucleon sources moving along forward (backward) rapidity, which contribute preferentially in the forward (backward) rapidity and from binary collision sources that contribute in a forward-backward symmetric manner,

\begin{eqnarray}
    \label{eq:dNchdeta_eta}
    \frac{dN_{Ch}}{d\eta} \left(\eta \right) &=&  \Big\{ \left(1-\textrm{x}_\textrm{hard}\right)\Big[\sum_{i=1}^{\textrm{N}_{\textrm{part}}^{A}} w_i\,N^{\textrm{sh},A}_{i,\textrm{part}}(\lambda)f_{+}\left(\eta\right) \notag\\
    &+& \sum_{i=1}^{\textrm{N}_{\textrm{part}}^{B}} w_i\,N^{\textrm{sh},B}_{i,\textrm{part}}(\lambda)f_{-}\left(\eta\right) \Big] \notag\\
    &+& \textrm{x}_\textrm{hard} \sum_{i=1}^{\textrm{N}_{\textrm{coll}}} w_i\,\textrm{N}^{\textrm{sh}}_{i,\textrm{coll}}(\lambda) \Big\} f\left(\eta\right)
\end{eqnarray}
\begin{equation}
    \label{f_eta_par}
    f\left(\eta_s\right) = exp\left( -\Theta \left( |\eta_s| - \eta_s^0 \right) \frac{\left(|\eta_s| - \eta_s^0\right)^2}{2\sigma^2} \right)
\end{equation}
\begin{equation}
    \label{fplusetaspace}
    f_+(\eta_{s}) =
    \begin{cases}
    0, & \eta_{s} < -\eta_T
    \\
    \frac{\eta_T + \eta_{s}}{2\eta_T}, & -\eta_T < \eta_{s} < \eta_T
    \\
    1, & \eta_{s} > \eta_T
    \end{cases}
\end{equation}
where $\textrm{N}_{\textrm{part}}^{A}$ and $\textrm{N}_{\textrm{part}}^{B}$ denote the numbers of participants from the forward- and backward-going nuclei, respectively, and the longitudinal profile satisfies $f_{-}(\eta_{s}) = f_{+}(-\eta_{s})$. The quantities $N^{\textrm{sh},A}_{i,\textrm{part}}(\lambda)$, $N^{\textrm{sh},B}_{i,\textrm{part}}(\lambda)$, and $\textrm{N}^{\textrm{sh}}_{i,\textrm{coll}}(\lambda)$ represent the shadow-suppressed contributions from the $i$th participant sources in the forward- and backward-going nuclei and from binary collision sources, respectively. While in the standard two-component MCGM, they are all unity, here accounting for the suppression due to the shadow effect, they turn out to be fractions less than unity. The extent of the suppression is controlled by the parameter $\lambda$ such that for $\lambda=0$, these are all unity and it represents the standard two-component MCGM. In this work, we take $\lambda=0.11$ as required to explain the ultra-central U+U data~\cite{Chatterjee:2015aja}. The parameters $\eta_{s}^{0}$ and $\eta_{T}$ control the plateau width and the tilt of the initial fireball, respectively, and $\Theta$ is the Heaviside step function. We set $\eta_{s}^{0}=1.3$ and $\sigma=1.5$, as taken from Ref.~\cite{Chatterjee:2017ahy}.


In this work, we use the FBMA to trigger the asymmetric events like the Body-Tip configurations. The FBMA is defined as the difference between charged particle multiplicities in the forward and backward pseudorapidity hemispheres~\cite{PHOBOS:2006mfc,He:2016qjs,ALICE:2015kal,Jia:2015jga,Sputowska:2020ghd}
\begin{equation}
\textrm{FBMA} =  \left| \int_{0}^{\eta_{\max}} \frac{dN_{\mathrm{ch}}}{d\eta} \, d\eta
- \int_{-\eta_{\max}}^{0} \frac{dN_{\mathrm{ch}}}{d\eta} \, d\eta \right|
\end{equation}
Unlike spectator-based measures, FBMA relies exclusively on charged particles, which are produced abundantly and detected with higher efficiency. In this study, we use $\eta_{\max}=1.0$, consistent with the STAR experimental acceptance.

\begin{figure}[t]
    \centering
    \includegraphics[width=0.46\textwidth]{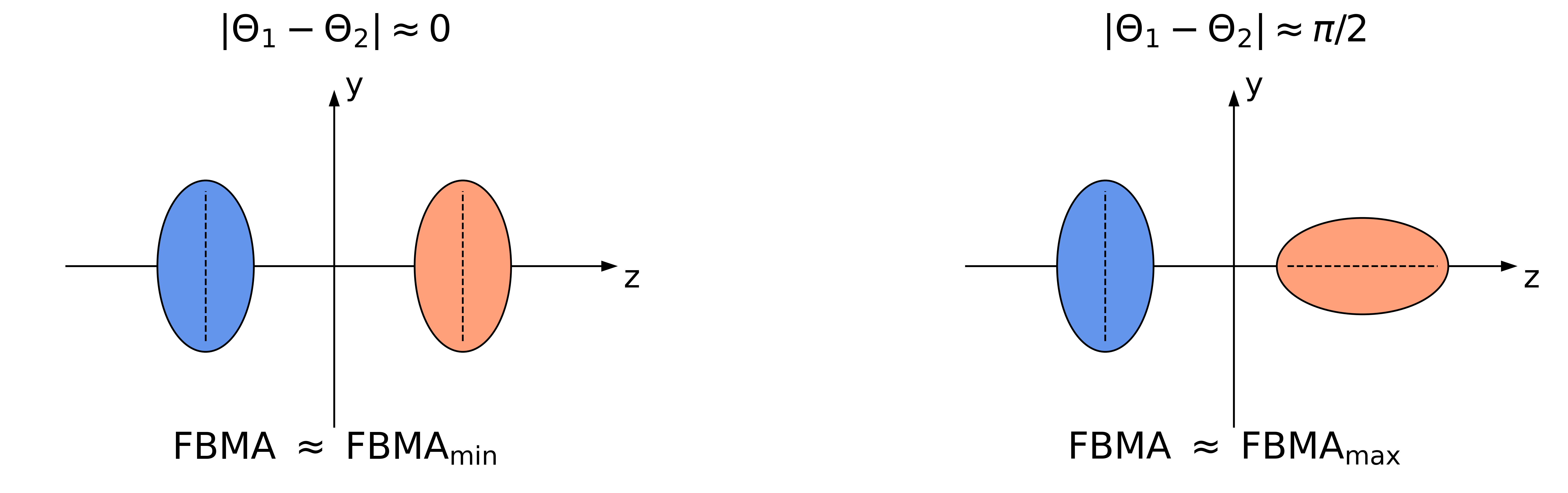}
    \par\bigskip
    \hspace*{0.1cm} $\Downarrow$ \hspace*{3.8cm} $\Downarrow$
    \par\bigskip
    \includegraphics[width=0.23\textwidth]{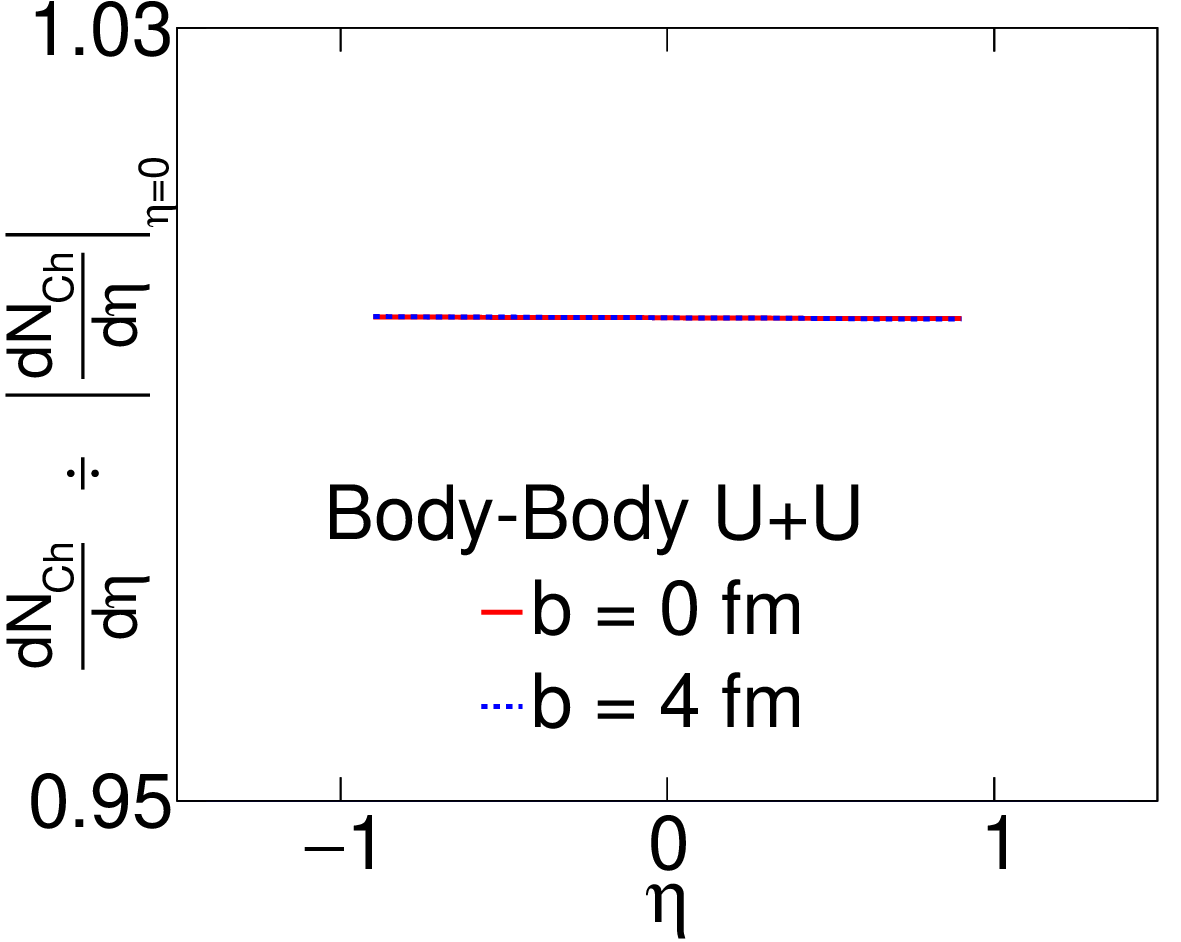}
    \includegraphics[width=0.23\textwidth]{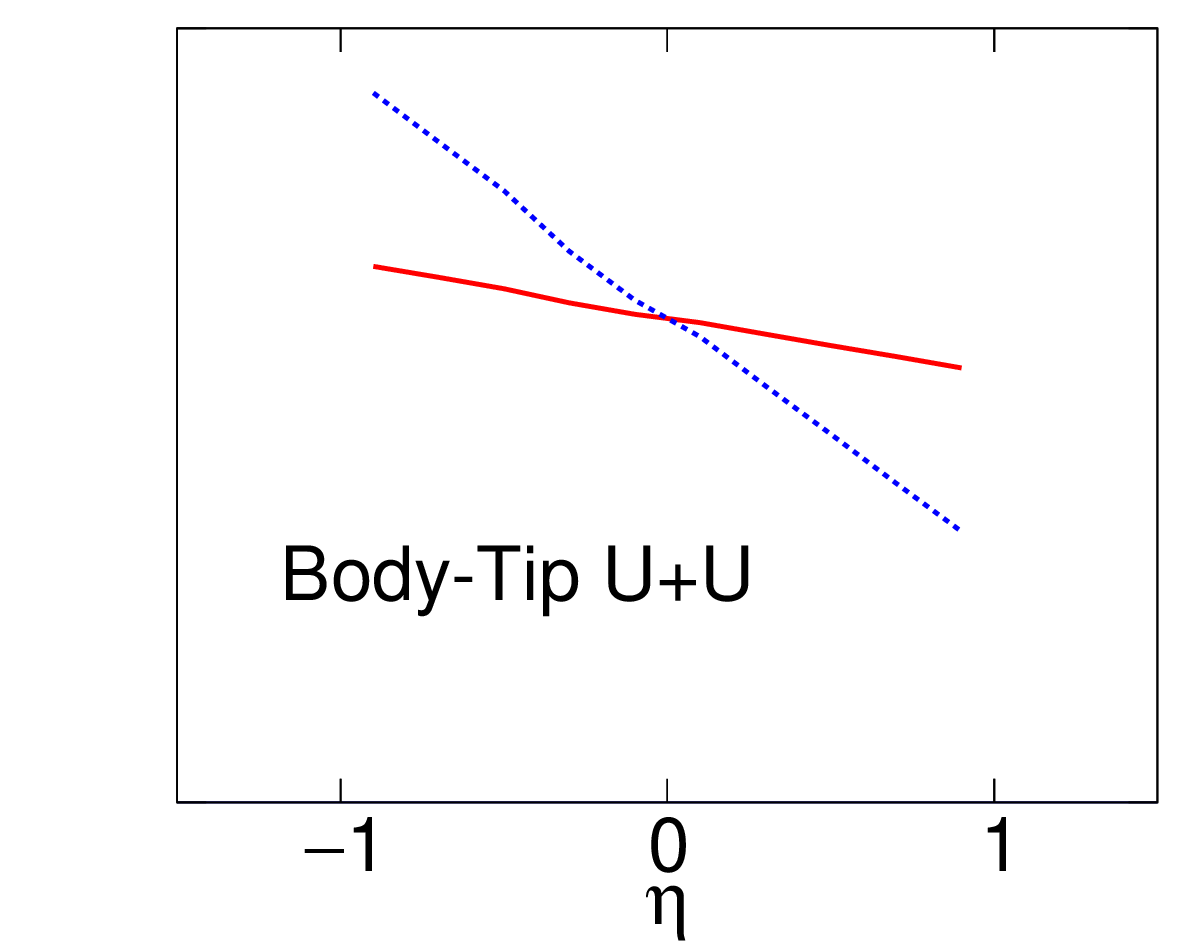}
    \caption{Top panels: Certain geometries of U+U collisions lead to an enhanced FBMA in the final state, arising from an increased relative angle between the major axes of the two nuclei at a fixed impact parameter. The beam direction is defined along the $z$-axis.\\
    Bottom panels: Pseudorapidity dependence of the charged-particle multiplicity scaled by its mid-rapidity value for the collision configurations shown, at impact parameters b = 0 and 4 fm.}
    \label{fig:FB_asymmetry}
\end{figure}

Figure~\ref{fig:FB_asymmetry} illustrates the dependence of the FBMA on collision geometry in U+U collisions. For symmetric orientations at fixed impact parameter, the asymmetry is dominated by event-by-event fluctuations in participant nucleon positions, with additional multiplicity fluctuations incorporated via a Negative Binomial convolution. The left panels show a Body--Body collision of two uranium nuclei, along with the resulting pseudorapidity distribution of charged particles scaled by its mid-rapidity value. As expected, this distribution is symmetric about mid-rapidity. As the relative orientation between the nuclei increases, geometric asymmetry in the overlap region becomes more pronounced, resulting in larger FBMA. The right panels show a Body--Tip collision, where one nucleus is aligned with the beam axis and the other is perpendicular. In this configuration the pseudorapidity distribution becomes markedly asymmetric about mid-rapidity, producing a significant FBMA. Due to the prolate deformation of uranium nucleus, Body--Tip collisions also reduce the initial spatial eccentricity of the overlap region. At fixed multiplicity, increasing FBMA correlates with decreasing initial eccentricity~\cite{Bairathi:2015uba}. Building on these observations, we introduce a method that uses FBMA as an event-selection handle to systematically vary the initial spatial eccentricity—and consequently elliptic flow—within a fixed centrality bin, while minimally affecting the magnetic field strength. This provides a pathway toward disentangling CME-driven charge separation from flow-induced backgrounds.

We employ Monte Carlo Glauber simulations incorporating realistic nuclear deformation and event-by-event fluctuations to quantify the relationship between FBMA and initial eccentricity in Au+Au and U+U collisions. Our results demonstrate that FBMA serves as an effective tuning parameter for controlling flow backgrounds while preserving sensitivity to CME-related observables. This approach complements existing techniques such as event-shape engineering~\cite{Schukraft:2012ah} and is readily applicable in current and future experimental analyses at RHIC and the LHC.

The use of charged particle FBMA thus offers a promising avenue for achieving a cleaner experimental separation of the CME from conventional background effects, advancing the study of local parity violation and topological phenomena in the QGP.

We generated 500 million events for each of the two collision systems: Au+Au and U+U. The Woods-Saxon parameters for U and deformed Au were obtained from Ref.~\cite{Shou:2014eya} and are listed in Table~\ref{table:woods-saxon_parameter}. The parameters for MCGM and shMCGM are determined by comparing the following observables with experimental data (for both Au+Au and U+U):\\
1. The variation of the probability distribution for multiplicity, $P(dN_{Ch}/d\eta)$~\cite{STAR:2015mki}.\\
2. The centrality dependence of $\textrm{N}_{\textrm{part}}$-scaled $dN_{Ch}/d\eta$~\cite{PHENIX:2015tbb}.\\
3. Multiplicity dependence of $v_2$~\cite{STAR:2015mki}.\\
4. $v_2-vs-dN_{Ch}/d\eta/\langle dN_{Ch}/d\eta \rangle$ in 1\% and 0.125\% ZDC events~\cite{STAR:2015mki}.

The model parameter values used in this study are listed in Table~\ref{table:model_parameter}. The value of the tilt parameter, $\eta_T$, for MCGM was taken from the references~\cite{Bialas:2004su, Chatterjee:2017ahy}. For shMCGM, we recalibrated $\eta_T$ by comparing the initial dipole asymmetry $\frac{\sum_i r^3\cos(\phi_i)}{\sum_i r_i^3}$ (summed over participant nucleon and binary collision positions) 
between MCGM and shMCGM.

\begin{table}[h!]
\centering
\caption{Parameter values for the Woods-Saxon distributions as used in this study.}
\label{table:woods-saxon_parameter}
\begin{tabular}{|c|c|c|c|c|}
\hline
System & R & a & $\beta_2$ & $\beta_4$\\
 \hline
Au & 6.37 & 0.535 & -0.13 & 0\\
 \hline
U & 6.81 & 0.55 & 0.28 & 0.093\\
 \hline
\end{tabular}
\end{table}

\begin{table}[h!]
\centering
\caption{Parameter values for the Glauber model as used in this study.}
\label{table:model_parameter}
\begin{tabular}{|c|c|c|c|c|c|}
\hline
Model & $\bar{n}$ & k & $\textrm{x}_\textrm{hard}$ &  $\eta_T$\\
 \hline
 MCGM & 2.1 & 1.1 & 0.14 & 3.33\\
 \hline
 shMCGM & 2.2 & 1.3 & 0.29 & 1.40\\
 \hline
\end{tabular}
\end{table}

\section{Results}
This section presents results on FBMA in Au+Au and U+U collisions, its correlation with forward-backward spectator neutron asymmetry (FBSA), and its utility as a control parameter for disentangling CME-like signals from flow-driven backgrounds.

\subsection{FBMA Distributions}
\begin{figure}[t]
\includegraphics[width=0.5\textwidth]{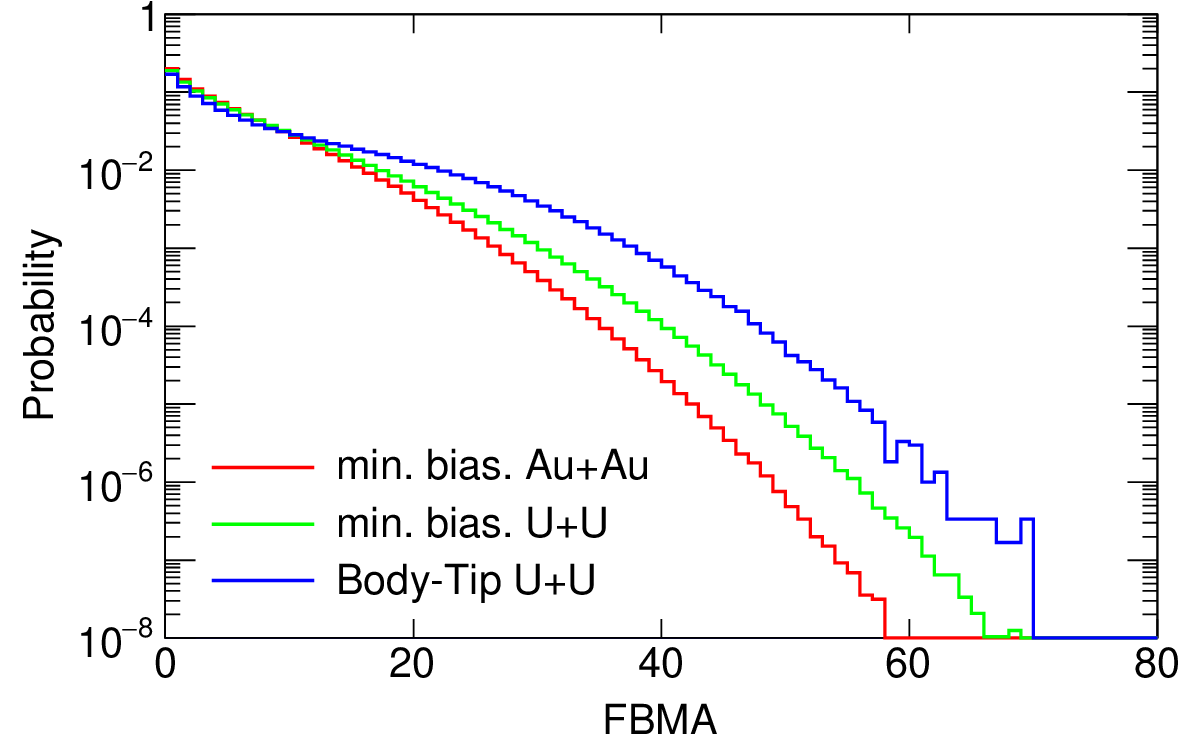}
\caption{\label{fig:mult_asymm_dists} Probability distribution of FBMA for minimum-bias Au+Au 200 GeV (red histogram) and U+U 193 GeV (blue histogram). The blue histogram with the largest asymmetry corresponds to U+U body–tip configurations.}
\end{figure}

Figure~\ref{fig:mult_asymm_dists} shows the probability distributions of FBMA for Au+Au collisions at $\sqrt{s_{NN}}=200$~GeV and U+U collisions at $\sqrt{s_{NN}}=193$~GeV. For U+U, results are shown both for minimum bias collisions and for the Body--Tip configuration, characterized by $\theta_1=\pi/2$, $\theta_2=0$, and $\phi_1=\phi_2=0$, which maximizes the asymmetry in participant nucleon numbers along the beam direction.

In Au+Au collisions, the FBMA distribution is relatively narrow. Since gold nuclei are nearly spherical with small oblate deformation, the FBMA arises predominantly from event-by-event fluctuations in nucleon positions, and the distribution is insensitive to nuclear orientation. Consequently, only minimum bias configuration for Au+Au results are shown.

In contrast, U+U collisions exhibit significantly broader FBMA distributions. The minimum bias U+U distribution extends to larger FBMA values than Au+Au, reflecting the impact of Uranium’s prolate deformation. The Body--Tip configuration yields the largest FBMA values, confirming that extreme forward--backward asymmetry is driven by highly asymmetric collision geometries. As a result, events with the largest FBMA in minimum bias U+U collisions are dominated by Body--Tip-like configurations. In the remainder of this work, U+U collisions refer to minimum bias events unless stated otherwise.

\subsection{Correlation Between FBMA and FBSA}
\begin{figure}[t]
\centering
\includegraphics[width=0.5\textwidth]{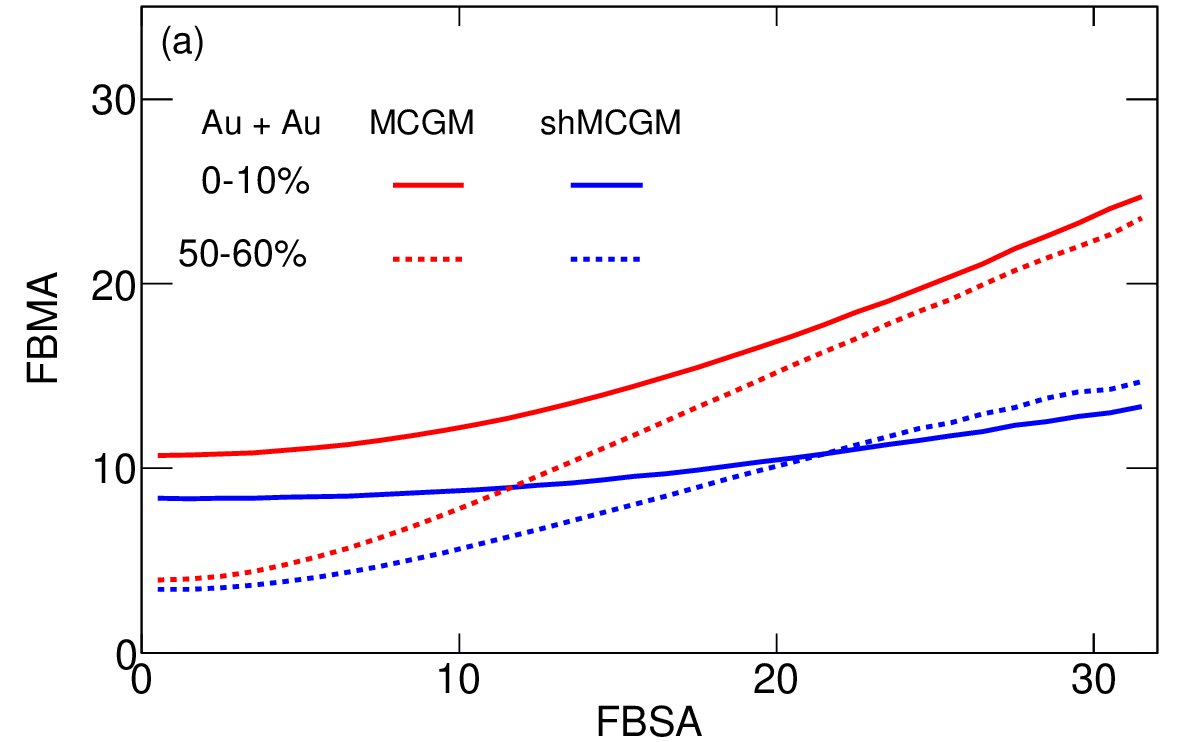}
\par\bigskip
\includegraphics[width=0.5\textwidth]{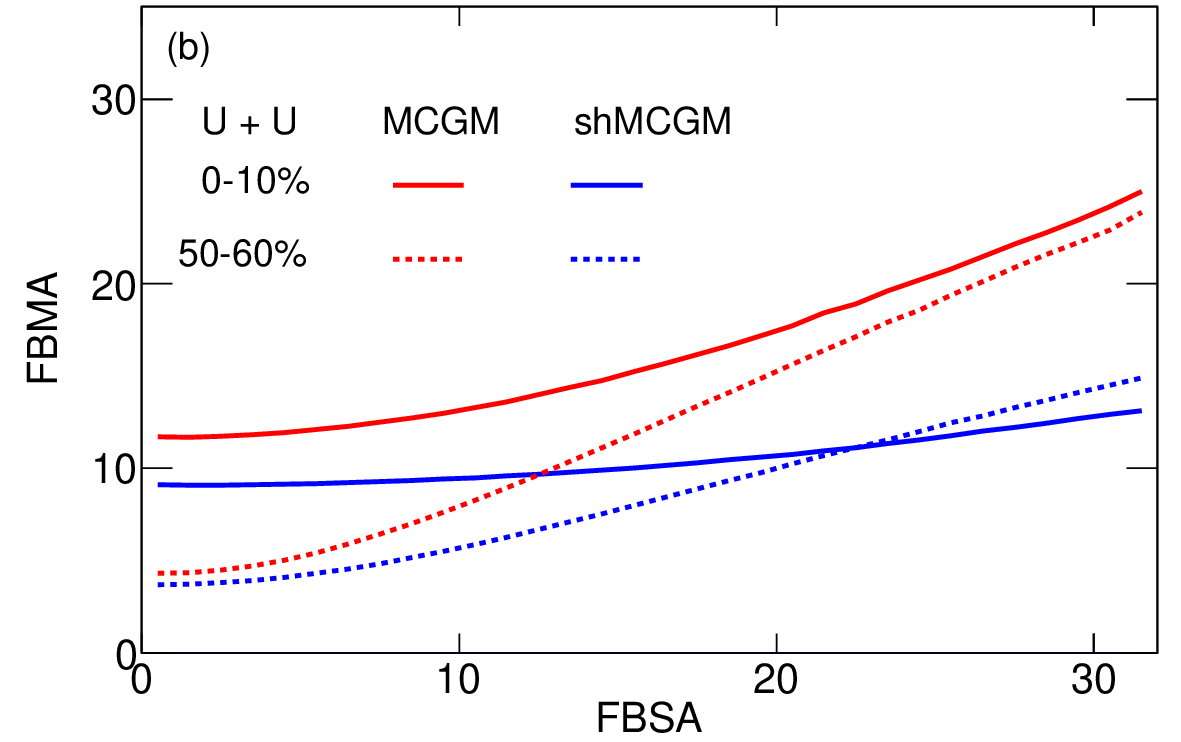}
\caption{\label{fig:AuAu_mult_asymm_vs_spec_asymm} FBMA versus FBSA for (a) Au+Au at 200 GeV and (b) U+U at 193 GeV. Results from MCGM (shMCGM) are shown by red (blue) lines. Solid and dashed lines denote 0–10\% and 50–60\% centrality bins, respectively.}
\end{figure}

The correlation between FBMA and FBSA is shown in Fig.~\ref{fig:AuAu_mult_asymm_vs_spec_asymm}: panel (a) presents Au+Au collisions at 200 GeV and panel (b) shows U+U collisions at 193 GeV, each for the 0--10\% and 50--60\% centrality classes. Results are compared between MCGM and shMCGM.

A positive correlation between FBMA and FBSA is observed for both systems and centralities, indicating that events with larger imbalance in spectator nucleons tend to produce a corresponding imbalance in charged particle production. This behavior reflects the underlying correlation between spectator geometry and participant matter distribution.

The correlation strength is systematically reduced in shMCGM relative to MCGM. While FBSA itself is unaffected by shadowing, nucleons experiencing stronger shadowing deposit less energy, leading to a suppressed FBMA for a given FBSA. The attenuation becomes more pronounced at larger FBSA values, where shadowing effects are stronger. When negative binomial fluctuations are removed, the FBMA--FBSA correlation becomes significantly tighter, confirming that multiplicity fluctuations partially obscure the underlying geometric correlation.

\subsection{Ellipticity and CME-like Correlators versus FBMA}
To investigate the interplay between elliptic geometry and CME-like observables, we compute the participant eccentricity,
\begin{equation}
\varepsilon_2 e^{i2\Psi_2^{\mathrm{PP}}}
= \frac{\sum_i r_i^2 e^{i2\phi_i}}{\sum_i r_i^2},
\end{equation}
where the sum runs over participant nucleon and binary collision positions. The participant plane angle $\Psi_2^{\mathrm{PP}}$ is rotated by $\pi/2$ to approximate the experimental event plane~\cite{Qin:2010pf}. For centralities up to $\sim$60\%, $\varepsilon_2$ serves as a reliable proxy for the elliptic flow coefficient $v_2$~\cite{Qiu:2011iv,Schenke:2013aza}.

CME-like effects are quantified using the magnetic field correlator~\cite{Bloczynski:2012en,Bloczynski:2013mca},
\begin{equation}
\gamma_B = e^2 B^2 \cos\left[2(\Psi_B-\Psi_{RP})\right],
\end{equation}
where $B$ and $\Psi_B$ are the magnitude and direction of the magnetic field evaluated at the center of the participant zone contributed by all protons at $t=0$ using Liénard--Wiechert potentials~\cite{Skokov:2009qp,Deng:2012pc}. In practice, $\Psi_{RP}$ is approximated by $\Psi_2^{\mathrm{PP}}$. This correlator is posited as a proxy for the CME signal $\gamma_{ab}$~\cite{Bloczynski:2012en, Bloczynski:2013mca}, drawing parallels to Eq.~\ref{eq:gamma_corr} of $\gamma_{ab}$ by substituting $\phi_a + \phi_b$ with 2$\Psi_B$.

\begin{figure}[t]
\centering
\includegraphics[width=0.5\textwidth]{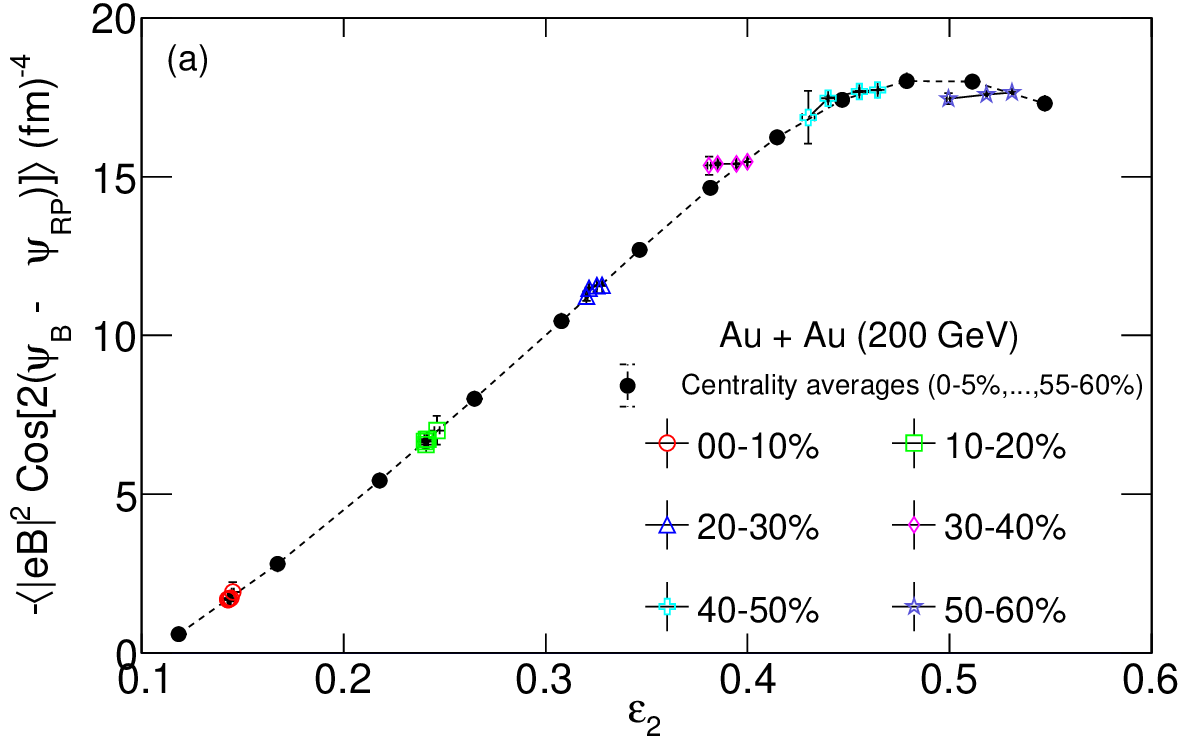}
\par\bigskip
\includegraphics[width=0.5\textwidth]{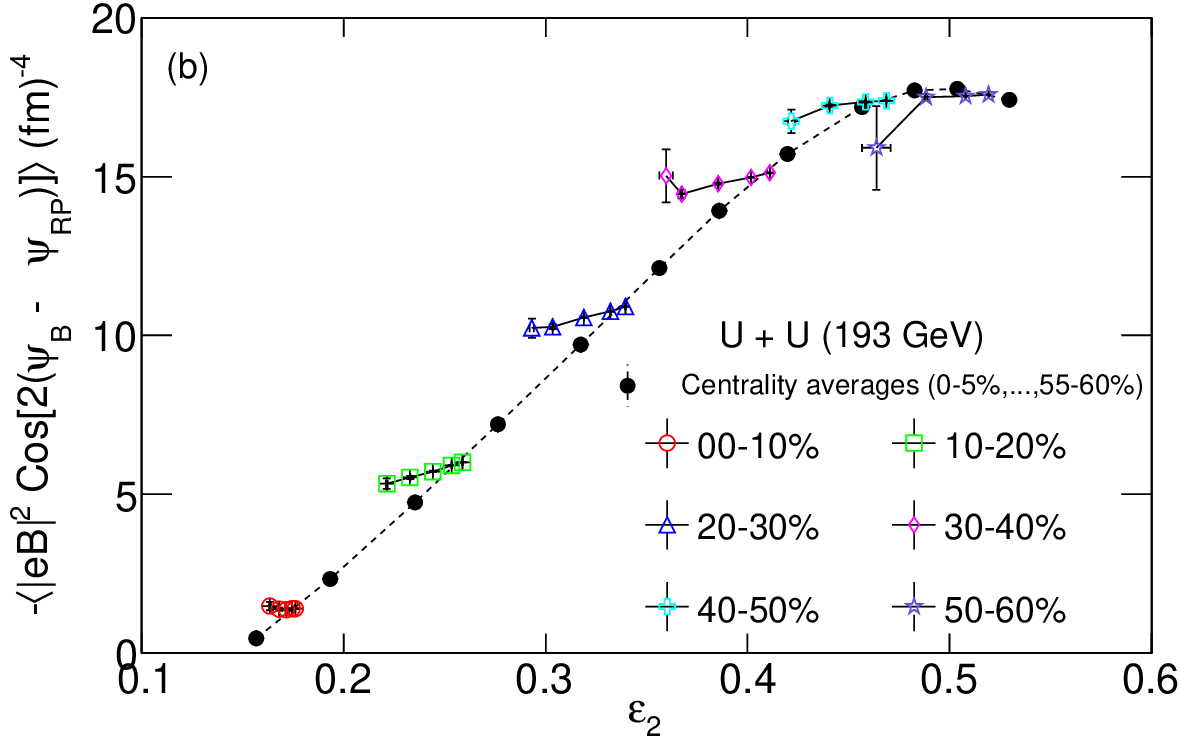}
\caption{\label{fig:CME_AuAu} Correlator $\gamma^{B}$ (with $\Psi_{RP}$ = $\Psi_2^{PP}$, the second order participant plane) at the center of the participant zone versus $\varepsilon_{2}$ for (a) Au+Au at 200 GeV and (b) U+U at 193 GeV, for various centrality and FBMA bins. Solid symbols connected by a black dashed line represent the FBMA = 0 bins for different centralities, illustrating a proportional increase of $\gamma_B$ with $\varepsilon_2$ across centrality.}
\end{figure}

Figure~\ref{fig:CME_AuAu} shows $\gamma_B$ versus $\varepsilon_2$: panel (a) presents Au+Au collisions and panel (b) shows U+U collisions, for different centrality and FBMA bins. For various centralities, the FBMA = 0 bins are portrayed through solid symbols with black dashed line, inferring a proportional augmentation of $\gamma_B$ with $\varepsilon_2$ across different centrality bins. This observation stipulates that merely altering centralities is insufficient to disambiguate CME from the collective response attributed to $\varepsilon_2$. However, in U+U an intriguing pattern emerges with the introduction of the FBMA parameter, facilitating the modulation of $\varepsilon_2$ within a specific centrality class. The disparate points within a centrality bin, delineated by open symbols, are derived when FBMA is varied within the range of 0 to maximum. For all instances, these varied curves emanate from the FBMA = 0 baseline, suggesting that a singular value of $\varepsilon_2$ corresponds to multiple values of $\gamma_B$ and vice-versa. In contrast, for Au+Au collisions, the introduction of additional binning in FBMA does not lead to a significant modification of the $\gamma_B$ correlator; within a given centrality class, the $\gamma_B$ values from different FBMA bins largely coincide with the centrality-averaged baseline, underscoring a qualitative difference from the behavior observed in U+U collisions. This phenomenon is particularly pronounced within the most central bins: (0-10)\% to (20-30)\%. This behavior suggests a strategy for experimental disentanglement: if the measured correlator $\gamma_{ab}$ is driven solely by flow-related backgrounds, then $\gamma_{ab}$--$v_2$ correlations obtained at fixed centrality should remain unchanged when FBMA is varied. In contrast, a genuine CME contribution would manifest as FBMA-dependent deviations in the correlation slope, analogous to those observed for $\gamma_B$ in Fig.~\ref{fig:CME_AuAu}.

These results demonstrate that FBMA provides an experimentally accessible and efficient handle on initial-state geometry, complementary to spectator-based selections using zero-degree calorimeters. FBMA-based event selection enables independent control over $\varepsilon_2$ and magnetic field strength, offering a promising avenue for isolating CME signals from flow-driven backgrounds.

\section{Conclusions}
The experimental identification of the CME in relativistic heavy-ion collisions remains challenging due to substantial backgrounds arising from collective flow. One promising strategy relies on the correlation between the magnetic field strength and the initial-state ellipticity, which in principle allows separation of CME-driven correlations from flow-induced backgrounds. However, previous suggestions based on spectator neutron asymmetry require highly efficient forward neutron detection, posing significant experimental difficulties~\cite{Chatterjee:2014sea}.

In this work, we have demonstrated that FBMA provides a robust and experimentally accessible alternative. In collisions involving deformed nuclei, such as uranium, the relative orientations of the nuclei naturally lead to configurations with unequal numbers of participating nucleons from the two colliding partners. This imbalance produces asymmetric energy deposition along forward and backward rapidities, resulting in a measurable FBMA. Since FBMA is constructed entirely from charged particle yields, it can be determined with sufficient efficiency using existing detector capabilities.

Using Monte Carlo Glauber simulations we showed that FBMA is strongly correlated with initial-state geometry and FBSA, while remaining sensitive to event-by-event fluctuations. In uranium--uranium collisions, large FBMA values are dominated by Body--Tip-like configurations, which exhibit reduced ellipticity due to the prolate nuclear shape. Exploiting this feature, FBMA can be varied within a fixed centrality class to modulate ellipticity independently of centrality and, to a large extent, magnetic field strength.

This additional degree of control enables a decisive test of CME-driven correlations. Within a given centrality, we find that a single value of ellipticity can correspond to multiple values of the magnetic-field correlator $\gamma_B$, and vice versa, when FBMA is varied. Notably, this effect exhibits a strong system dependence: while in U+U collisions the variation of FBMA leads to a significant spread in $\gamma_B$ at fixed $\varepsilon_2$, in Au+Au collisions the $\gamma_B$ values from different FBMA bins largely coincide with the centrality-averaged baseline, indicating a much weaker sensitivity. This qualitative difference highlights the enhanced discriminating power of FBMA in deformed systems. Such behavior is not expected if charge-dependent correlations are driven solely by flow, but is consistent with the presence of a CME contribution. Therefore, FBMA-based event selection offers a clear pathway to disentangle CME signals from flow-related backgrounds.

In summary, FBMA emerges as a powerful and practical tool for CME studies. When combined with conventional centrality and event-shape selection techniques, FBMA enhances sensitivity to initial-state magnetic-field effects while avoiding the experimental limitations associated with spectator neutron measurements. This approach provides a promising avenue for future experimental efforts aimed at establishing unambiguous evidence for CME and related topological phenomena in the quark--gluon plasma, particularly by leveraging the enhanced sensitivity observed in collisions of deformed nuclei such as U+U compared to nearly spherical systems like Au+Au.


\newpage
\bibliography{main}

\end{document}